# Molecular characterization of wild Pleurotus ostreatus (MW457626) and evaluation of β-glucans polysaccharide activities

Ghazwan Q. Hasan Ph.D.
*Department of Biology, College of Education for Pure Sciences, University of Mosul, Iraq*, ghazwan.qasim@yahoo.com

Shimal Y. Abdulhadi Ph.D.
*Department of Biology, College of Education for Pure Sciences, University of Mosul, Iraq*

# Molecular characterization of wild Pleurotus ostreatus (MW457626) and evaluation of **β**-glucans polysaccharide activities

## Abstract

*Pleurotus ostreatus* is a common cultivated edible mushroom worldwide. The fruiting bodies of *P. ostreatus* is a rich source of a *β*-glucans polysaccharide. The current study aimed to investigate the effectiveness of *β*-glucans as a natural polysaccharide produced by *P. ostreatus* as an antioxidant, antimicrobial, and anticancer. The molecular identification of *P. ostreatus* isolate was confirmed by Internal Transcribed Spacer (ITS) sequence. The sequence alignment and phylogenetic evolutionary relationship of studied ITS sequence were performed against some deposited sequences in GenBank. The analysis of high-performance liquid chromatography (HPLC) as well as the result of fourier transform infrared spectroscopy (FTIR) has confirmed the presence of *β*-glucans polysaccharide in the tested samples. The percentage of antioxidant activity of *β*-glucans showed a gradual increase from 8.59% to 12.36, 18.56, 23.69, 44.66 and 80.36% at the concentrations of 31.2, 64.4, 125, 250, 500, and 800 µg/ml, respectively. In addition, all concentrations of *β*-glucans showed higher antioxidant activities when compared with standard antioxidant (Vitamin C). The highest- antimicrobial activity of *β*-glucans polysaccharide was against *P. aeruginosa* with a zone of inhibition (45 mm), while the lowest activity was against *S. aureus* (13 mm) both at 100 mg/mL. The percentage of growth-inhibiting of MCF-7 a human breast cancer cell line and normal WRL-68 cell line affected by *β*-glucans were determined by 3-(4,5)-dimethylthiazol (-z-y1)-3,5-di-phenytetrazoliumromide (MTT assay). The results revealed that the percentage of growth inhibiting of MCF-7 and WRL-68 cells were gradually increased in both lines and was the highest with MCF-7 line, where the percentage were 18, 23, 50, 59, and 62 % compared with WRL-68 line which showed 4, 6, 9, 13, and 22% at 1.0, 1.5, 2.0, 2.5, and 3.0 µg/ml, respectively. This present paper revealed that *P. ostreatus* has therapeutic values that can be used as a natural medicine instead of industrial compounds.

RESEARCH PAPER

# Molecular Characterization of Wild *Pleurotus Ostreatus* (MW457626) and Evaluation of β-glucans Polysaccharide Activities

Ghazwan Q. Hasan*, Shimal Y. Abdulhadi

Department of Biology, College of Education for Pure Sciences, University of Mosul, Iraq

**Abstract**

***Pleurotus ostreatus*** **is a common cultivated edible mushroom worldwide. The fruiting bodies of *P. ostreatus* is a rich source of a β-glucans polysaccharide. The current study aimed to investigate the effectiveness of β-glucans as a natural polysaccharide produced by *P. ostreatus* as an antioxidant, antimicrobial, and anticancer. The molecular identification of *P. ostreatus* isolate was confirmed by Internal Transcribed Spacer (ITS) sequence. The sequence alignment and phylogenetic evolutionary relationship of studied ITS sequence were performed against some deposited sequences in GenBank. The analysis of high-performance liquid chromatography (HPLC) as well as the result of fourier transform infrared spectroscopy (FTIR) has confirmed the presence of β-glucans polysaccharide in the tested samples. The percentage of antioxidant activity of β-glucans showed a gradual increase from 8.59% to 12.36, 18.56, 23.69, 44.66 and 80.36% at the concentrations of 31.2, 64.4, 125, 250, 500, and 800 μg/ml, respectively. In addition, all concentrations of β-glucans showed higher antioxidant activities when compared with standard antioxidant (Vitamin C). The highest-antimicrobial activity of β-glucans polysaccharide was against *P. aeruginosa* with a zone of inhibition (45 mm), while the lowest activity was against *S. aureus* (13 mm) both at 100 mg/mL. The percentage of growth-inhibiting of MCF-7 a human breast cancer cell line and normal WRL-68 cell line affected by β-glucans were determined by 3-(4,5)-dimethylthiazol (-z-y1)-3,5-di-phenytetrazoliumromide (MTT assay). The results revealed that the percentage of growth inhibiting of MCF-7 and WRL-68 cells were gradually increased in both lines and was the highest with MCF-7 line, where the percentage were 18, 23, 50, 59, and 62% compared with WRL-68 line which showed 4, 6, 9, 13, and 22% at 1.0, 1.5, 2.0, 2.5, and 3.0 μg/ml, respectively. This present paper revealed that *P. ostreatus* has therapeutic values that can be used as a natural medicine instead of industrial compounds.**

## 1. Introduction

Mushrooms have been previously played a crucial role in folk medicine [1], and utilized in popular medicine of China and some Asian countries. They have antioxidant, antibacterial anticancer, antiviral, anti-inflammatory, hypoglycemic, and hypocholesterolemic characteristics [2–4]. Nowadays, edible mushrooms are popular beneficial foods with low calories, carbohydrates, fat and cholesterol-free [5]. Mushrooms contain some metal elements and vitamins such as calcium, potassium, phosphorus, copper, ascorbic acid, vitamin D and K, niacin, thiamine and others [6]. The content of macro and micronutrients in *Pleurotus* spp. varies according to the growth substrate [7]. The content of macro and micronutrients in *Pleurotus* spp. varies according to the growth substrate. For instance, Ca and K content in *Pleurotus* spp. were ranged from 1 to 330, and from 271 to 4054.3 mg/100 g d.w. respectively [7]. Moreover, vitamin D2 content in *Pleurotus* spp is low, but *Pleurotus* spp. have a high content of ergosterol, which generates vitamin D2 upon UV light exposure [8] They are rich also in some secondary metabolites which have high medicinal value such as polysaccharide, lecithin, phenolic compounds, terpenoids, saponin, polyketides, and others [9,10]. Several studies revealed that the *Pleurotus* genus appears multidirectional benefits for supporting human health [11–13]. Further, the oyster mushroom was considered a functional food, due to its positive role in a human being [14,15]. The *Pleurotus* genus consists of around 14 species [16]. Mushrooms are enriched with minerals, proteins, and vitamins [17]. The therapeutic properties and bioactive compounds were found in both fruiting bodies and the mycelia of oyster mushrooms [18]. There are two different compounds found in *Pleurotus* mushrooms according to their molecular weight, high-molecular weight secondary metabolites which mainly consist of polysaccharides, involving $\beta$-glucans, proteins and peptides; and low -molecular weight secondary metabolites involve polyphenols, terpenes and fatty acids [19]. The outstanding compounds in *Pleurotus* mushrooms are the $\beta$-glucans polysaccharides which are mainly found in the cell wall of mushrooms. It involves glucopyranose units binding with glycosidic bonds at the positions of (1 $\rightarrow$ 3)-$\beta$, (1 $\rightarrow$ 6)-$\beta$- or (1 $\rightarrow$ 3)-$\alpha$. $\beta$-glucans. It has many important roles such as activation of the human immune system, antioxidant, antimicrobial, anticancer, antiviral, antifungal, controlling the cholesterol level and glucose regulation in blood [20]. The *G. lucidum* polysaccharides have also similar activity of anticancer [21,22]. $\beta$–glucans inhibits the growth of cancer cells by stimulating the immune reaction in the active-cells [23]. It was found that $\beta$-glucans encourages the inducing of interferon-gamma and interleukin-12 (IL-12) in lymphocytes [24] and increasing of T helper 1 response through inducing of interleukin-1 (IL-1), nitric oxide and tumor necrosis factor-alpha [25]. Oyster mushroom is one of many richest sources of $\beta$-glucans polysaccharide [26]. Based on [27] the total glucans in *P. ostreatus* is 25.636 g/100 g dm (dry mass), while the $\beta$-glucans concentration is (24.230 g/100g dm). However, the total concentrations of $\beta$-glucose were (4.6 g/100 g dm and 9 g/100 g dm) in mycelia and fruiting bodies of *P. ostreatus*, respectively [28]. FTIR technique helps to detect the active groups that appear in the sample and is commonly used to examine the fungi and plant-based polysaccharides [22,29]. The ITS region is widely considered to be the most important sequenced DNA region used for molecular detection purposes in fungi [30,31] and is considered the universal barcode sequence in fungi [32]. The purposes of the present study are to perform some molecular analysis such as multiple sequence alignments and draw a phylogenetic tree of studied ITS sequence with deposited sequences in GenBank, and study the antioxidant, antimicrobial, and anticancer activities The macroscopic features of the Basidiomycota were recorded, based on the approach followed by Ref. [34]. The collected isolates were deposited at the Herbário do Estado Maria Eneyda P. Kaufmann Fidalgo (SP) under the following names CCB001, Pt II, Pt III, and Pt IV and their accession numbers were SP392848, SP392847, SP392849, and SP392850, respectively. The microscopic analysis involved rehydrates the dry material with 70% ethanol, then with 5% KOH reagent [35]. For more morphological characterizations, a consultation was needed from Refs. [36–45].

* Corresponding author at:
E-mail addresses: dr.ghazwan@uomosul.edu.iq, ghazwan.qasim@yahoo.com (G.Q. Hasan).

## 2. Materials and methods

### 2.1. Sampling and isolation of P. ostreatus

*P. ostreatus* fruiting bodies growing on the trunks of Morus tree (Fig. 1) have been collected from the farms located in Rashidiyah region/Mosul city, Iraq https://goo.gl/maps/tFJwGPjRHrPWbUFD8. Samples were placed in sterile polyethylene bags and transferred to the lab, then washed with running water several times, after which they were cut lengthwise into small parts in thin slices. The pieces were sterilized with 70% ethyl alcohol for 2 min, then washed with sterile distilled water for 2–5 minutes, after which the samples were left to dry on sterile filter papers Whatman No. 1, then the pieces were distributed on Petri dishes containing Potato Dextrose Agar using sterile forceps with alcoholic flame. The dishes were covered with aluminium foil and transferred to the incubator at 28 °C for 7–10 days. Growing mycelia were preserved in a suitable place until use [33].

### 2.2. Identification

#### 2.2.1. Phenotypic identification

The macroscopic features of the Basidiomycota were recorded, based on the approach followed by Ref. [34]. The collected isolates were deposited at the Herbário do Estado Maria Eneyda P. Kaufmann Fidalgo (SP) under the following names CCB001, Pt II, Pt III, and Pt IV and their accession numbers were SP392848, SP392847, SP392849, and SP392850, respectively. The microscopic analysis involved rehydrates the dry material with 70% ethanol, then with 5% KOH reagent [35]. For more morphological characterizations, a consultation was needed from Refs. [36–45].

### 2.2.2. Molecular identification

Genomic DNA from pure mycelia of *P. ostreatus* was isolated using the CTAB method as described by Ref. [46]. The DNA fragment of ITS was amplified by PCR using ITS1 forward (5′- TCCGTAGGTGAACCTGCGG -3′) and ITS4 reverse (5′- TCCTCCGCTTATTGATATGC-3′) primers. The DNA fragment of 657 bp was checked by electrophoresis on a 1% agarose gel and visualized by a UV transilluminator (Fig. 2). The PCR product was sent for sequencing to Microgen Company/South Korea https://dna.macrogen.com/#. The ITS sequence analysis of *P. ostreatus* was performed with the BLASTn site at the NCBI https://blast.ncbi.nlm.nih.gov/Blast.cgi.

## 2.3. Extraction of β-glucans

The *β-glucans* polysaccharide was extracted from *P. ostreatus* using the approach presented by Ref. [47].The cultured mycelia were extracted by hot water for 60 min, the culture was then filtered with Whatman No. 1 filter paper. The samples were precipitated with 4 vol of 95% (v/v) ethanol and left for a day at 4 °C. Next, the centrifugation of prepecipitates was performed at 10 000×*g* for 10 min, and the supernatant was discarded. The precipitates were lyophilized into powders. The powders then were redissolved in deionized water, and the final stock concentration was 20 mg/ml, and then kept at 4 °C until use.

## 2.4. β-glucans activities

### 2.4.1. Measurement of antioxidant efficacy of β-glucans

The purified β-glucans isolated from *P. ostreatus* were subjected to antioxidant measurement using 2,2-diphenyl-1-picrylhydrazyl or DPPH radical scavenging technique [48]. The stock concentration of 400 μg/ml was made by dissolving 0.04 g of DPPH into 100 ml of methanol. The control solution was made by dissolving 0.5 g of ascorbic acid into 100 ml of methanol and dH2O in a 1:1 ratio. The concentrations (31.2, 64.4, 123, 250, 500 and 800 μg/ml) of β-glucans and ascorbic acid were prepared (Fig. 6). 1 ml of DPPH was mixed with 1 ml of each prepared concentration of β-glucans and ascorbic acid, the mixture then was vigorously shaken and left in dark at room temperature for a half an hour. The control solution contained the same volume as the reaction solution except using methanol solvent instead of the extracts above. The absorbance of DPPH was determined at the wavelength 517 nm. The DPPH value was expressed as a percentage, according to the method used by Ref. [49] as in the following equation:

$$\text{Percentage of scavenging} = A^{O} - A1 / A^{O} \times 100$$

$A^{O}$ = *Absorbance of control;*

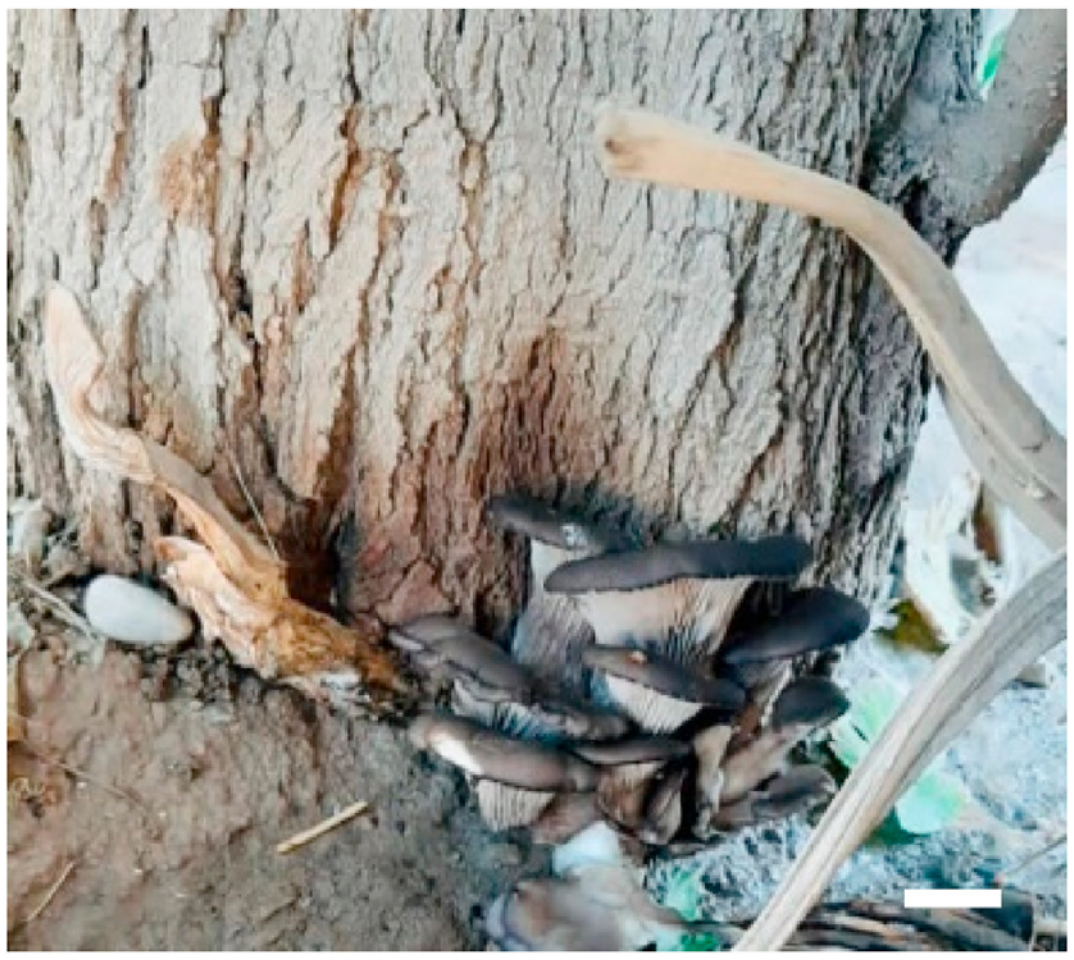

*Fig. 1. Morus tree carrying fruiting bodies of P. ostreatus. Scale bar: 1 cm.*

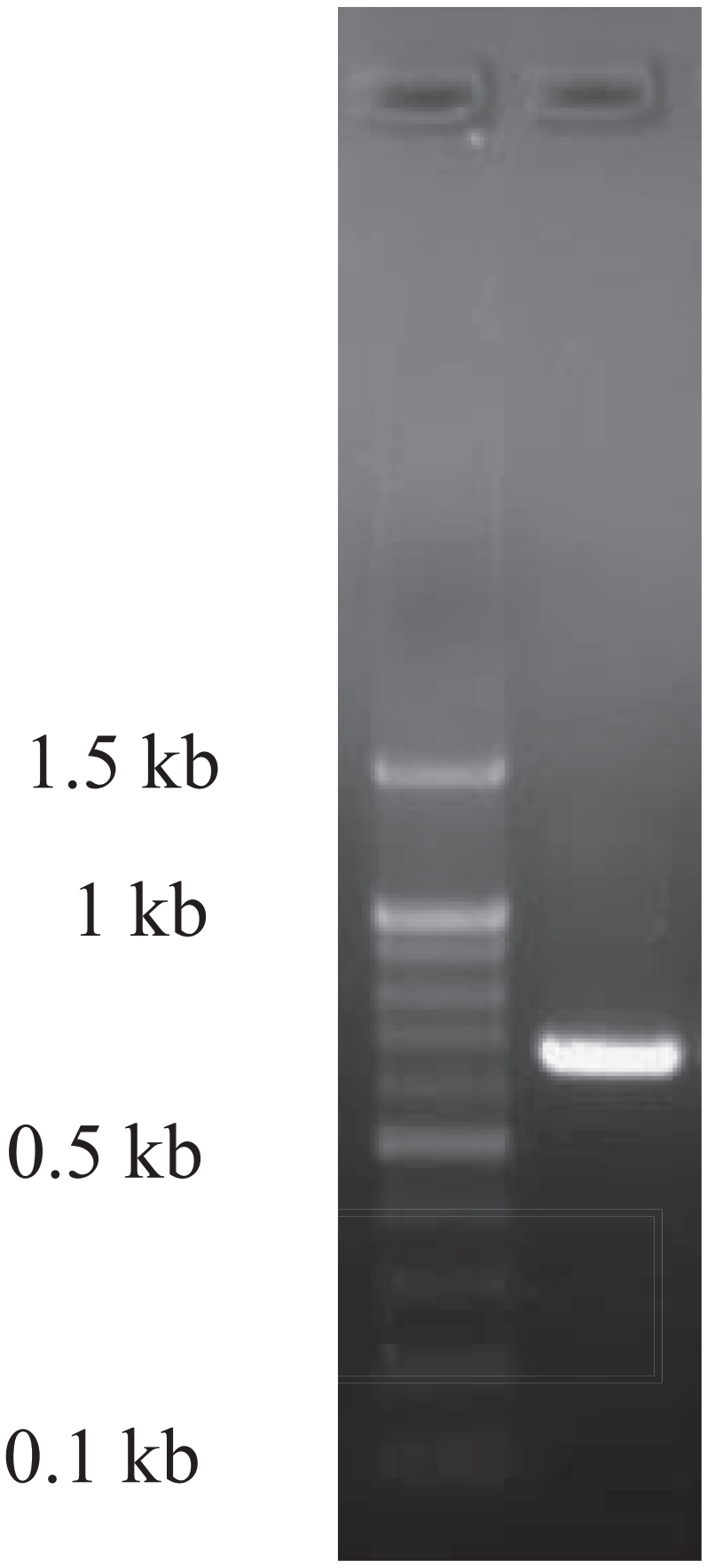

*Fig. 2. Agarose gel image of amplified ITS fragment (657 bp) of P. ostreatus. (1) 100 bp DNA ladder, (2) ITS fragment amplified with ITS1 forward and ITS4 reverse primers.*

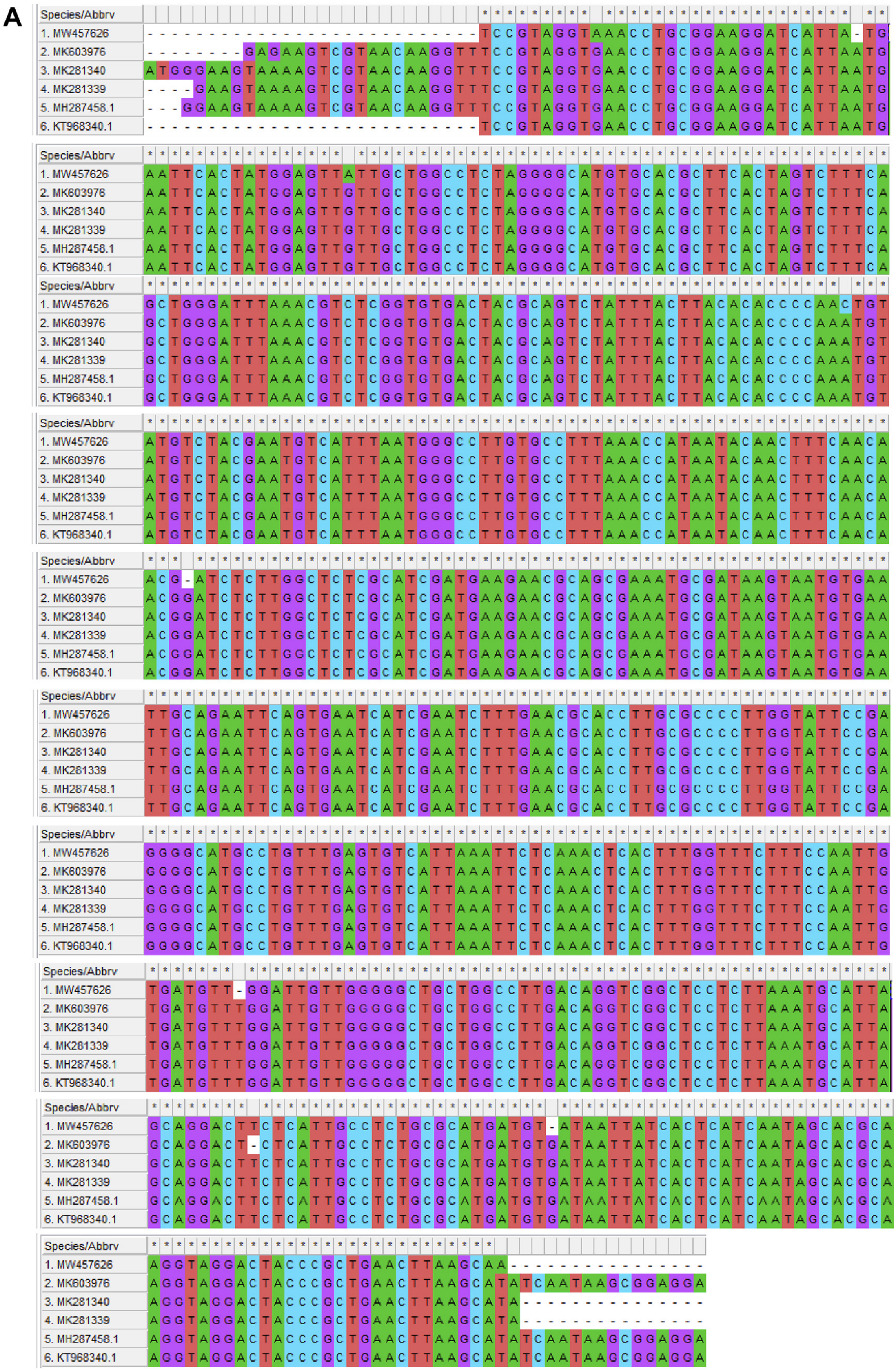

*Fig. 3. (A) Multiple sequence alignments of ITS region of P. ostreatus (Accession: MW457626, distinguished by red rectangle) with some deposited sequences in GenBank (Accessions: MK603976, MK281340, MK281339, MH287458.1 and KT968340.1). (B) The phylogenetic tree was deduced by the UPGMA approach [53]. The associated taxa clustered together at bootstrap (500 replications) are appeared next to the branches, and represented as percent of replicate trees [54]. The phylogenetic tree was drawn with the same branch length units used in the evolutionary distances. These distances were figured out using the Maximum Composite Likelihood method [55] and are in the units of the number of base substitutions per site. This analysis consisted of 6 nucleotide sequences. All ambiguous positions were excluded for each sequence pair. The total positions were 705 in the final dataset. Evolutionary analysis was applied in MEGA X_2.2 [56].*

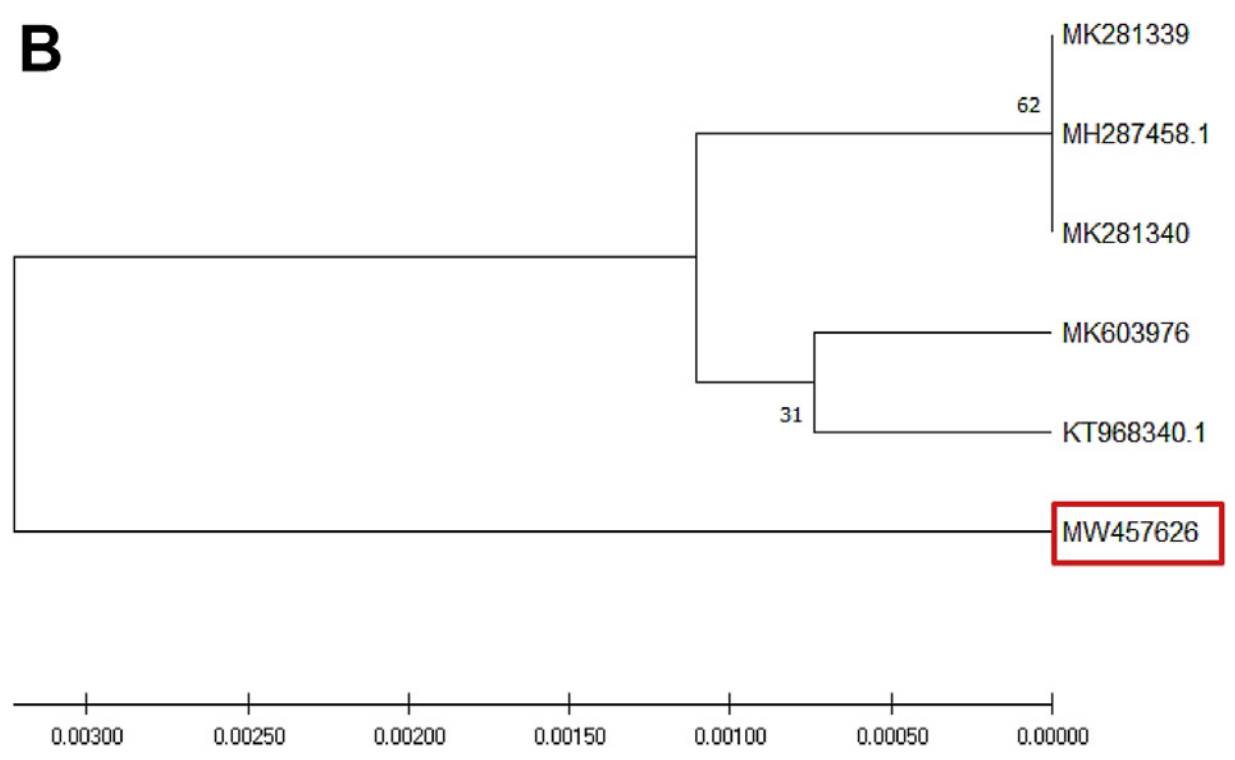

*Fig. 3. (Continued).*

*A1 = Absorbance of sample.*

*2.4.2. Assessment of antimicrobial activity of β-glucans*

The antimicrobial properties of β-glucans were accomplished with the agar well diffusion method [50]. The β-glucans extract was dissolved in Dimethyl sulfoxide of 100 mg/mL, the chloramphenicol at 40 mg/mL and dH2O were used as positive and negative controls, respectively. The selective wells poured into Petri dishes (5-mm in diameter) containing 2% nutrient agar (2% agar, 0.5% Peptone, 0.5% sodium chloride, 0.3% beef extract, 1 L distilled water, at pH 7.0). The test was accomplished in triplicate, and the cultures were kept overnight at 28 °C. For antimicrobial activity, the plates were incubated at convenient growth temperatures for tested bacteria (*Pseudomonas aeruginosa*, *Escherichia coli*, *Staphylococcus aureus*, and *Proteus mirabilis*). The diameter of the inhibition zone was measured in mm [51] as shown in Fig. 7 and Table 3.

*2.4.3. Determination of cytotoxic activity of β-glucans*

MCF-7 a human breast cancer cell line and WRL-68 a normal cell line were kindly provided from the Biotechnology Research Center of Al-Nahrain University/Baghdad and used to determine the anticancer activity of β-glucans (Table 4). The MTT assay is a proliferation detection and colorimetric cytotoxicity test, this approach is depending on the metabolic effects of normal cells [52]. The tetrazolium salts are usually decreased in variable cells to a blue-coloured formazan, which is proportional to the number of active cells. The MCF-7 and WRL-68 cell lines ($5 \times 10^5$ cells/mL) were placed into a 96-well microplate, the cultures were grown overnight at standard conditions of 5% $CO_2$ ml at 37 °C. The cells were incubated for 48 hours with different concentrations of β-glucans (1.0, 1.5, 2.0, 2.5, and 3.0 μg/ml) (Table 4). Each sample was prepared with 1% of fetal bovine serum (FCS) medium. Following incubation, 1 mg/mL of MTT was added to each sample and incubated once again for 3 h. Then, 100 μL of the sodium dodecyl sulfate solution were applied to each sample to dissolve formazan crystals, and the wells were incubated overnight at 37 °C. The optical density was determined at 570 nm, and then the cytotoxicity of the cancer cell lines was calculated as a percentage based on (100% of living cells in the control).

## 3. Results

*3.1. Isolation of genomic DNA, PCR amplification, and ITS sequence analysis*

Genomic DNA from fruiting bodies of *P. ostreatus* was extracted using the CTAB method. The ITS fragment 657 bp was successfully amplified by PCR using forward (ITS1) and reverse (ITS4) primers and clearly shown on gel electrophoresis (Fig. 2). The PCR product was sent for sequencing to Microgen Company/South Korea https://dna.macrogen.com/#. The ITS sequence was collected in high quality and then deposited in GenBank under accession number MW457626 which showed a homology of 99% with some deposited sequences in GenBank (Accessions: MK603976, MK281340, MK281339, MH287458.1 and KT968340.1). Multiple sequence alignments and evolutionary relationships were performed for the ITS region by MEGA X_10.2.2 software https://www.megasoftware.net/. (Fig. 3: A and B).

*3.2. HPLC analysis of β-glucans polysaccharide*

1 mL of extract was analyzed by HPLC [57]. This analysis was accomplished in the Department of Environment and Water/Ministry of Science and Technology, following the acid hydrolysis process using the HPLC device (SYKAM model-Germany). The used column was C18-ODS with dimensions 4.6 mm* 25 cm. The measurements were taken by (UV–Vis) at 360 nm wavelength with a flow rate 1 mL per minute. The analysis revealed the presence of β-glucans polysaccharide as a main bioactive compound (Fig. 4 and Table 1).

*3.3. Fourier-transform infrared spectroscopy analysis of β-glucans polysaccharide*

FTIR is an assay used to collect an infrared spectrum of absorption or emission from different states of matter (solid, liquid or gas). FTIR spectrum analysis pointed to the existence of β-glycosidic bonds. The results revealed the presence of an NH group of frequency 3344 $cm^{-1}$, =C–H group at the frequency of 3078 $cm^{-1}$, C–H group at 2899 $cm^{-1}$, also the presence of C=O, N–O, C–H and C–N groups at the frequencies of (1685, 1564, 1413 and 1066 $cm^{-1}$) respectively. However, the frequencies that are less than 1000 $cm^{-1}$ belong to the heavy elements. The data are shown in Fig. 5 and Table 2.

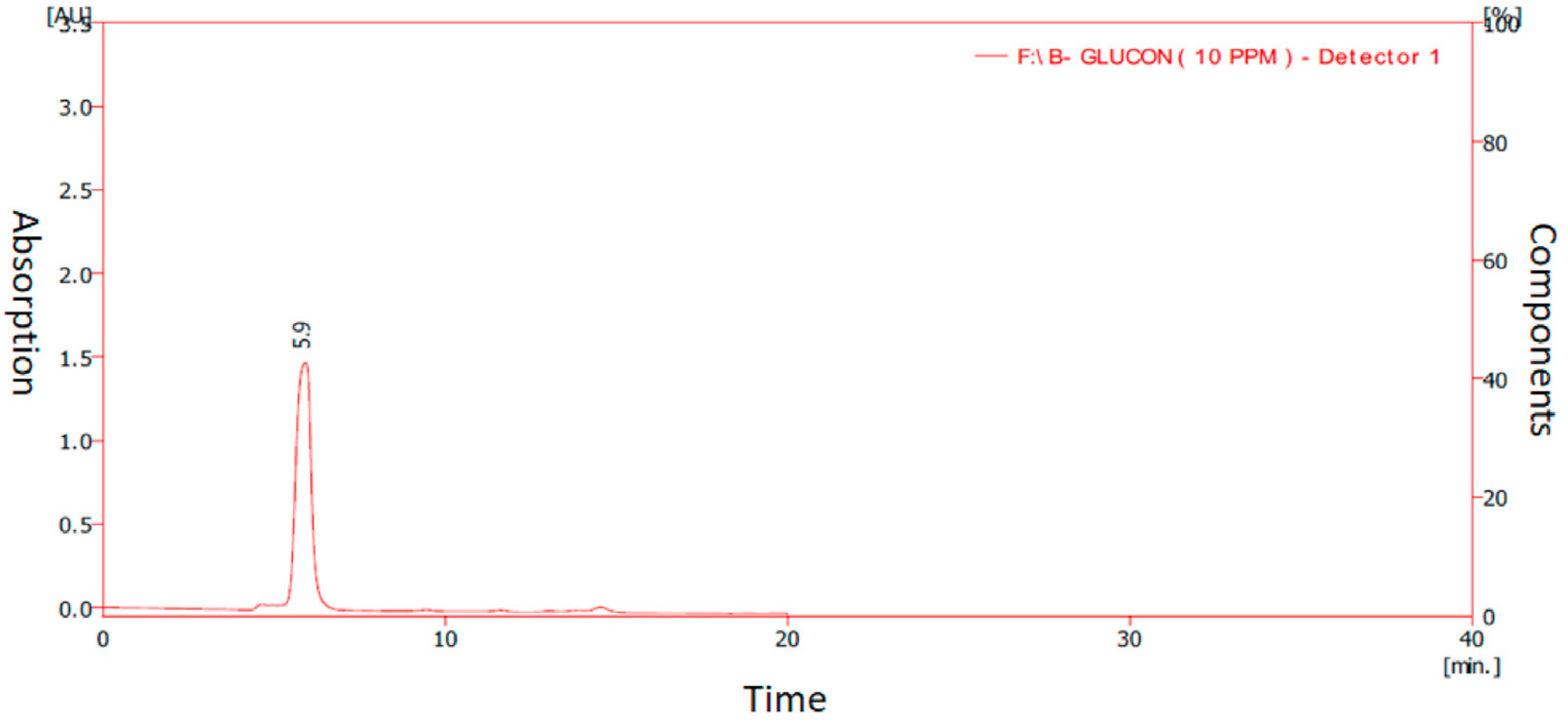

*Fig. 4. HPLC chromatogram of polysaccharide β-glucans extracted from P. ostreatus. The samples were detected by (UV–Vis) at a wavelength of 360 nm at a flow rate 1 mL per minute.*

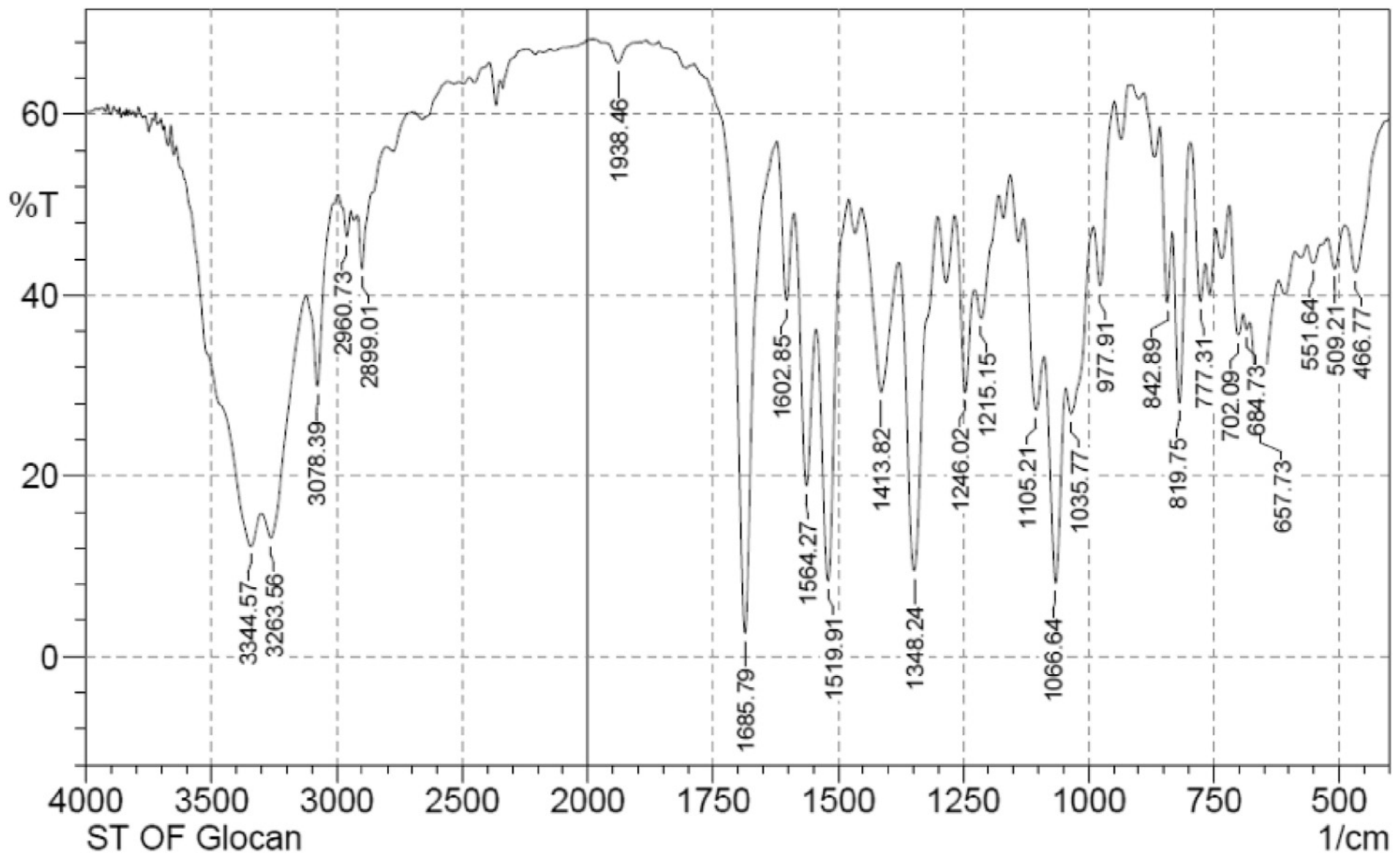

*Fig. 5. FTIR spectra of β-glucans extracted from P. ostreatus fruiting bodies.*

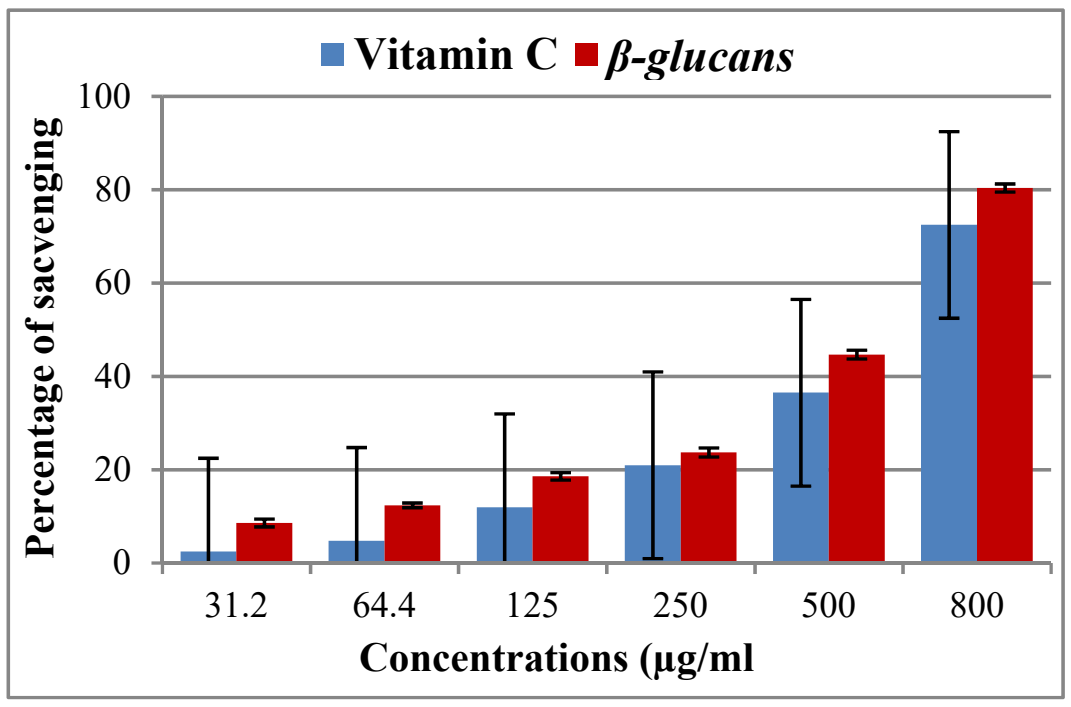

*Fig. 6. Percent of DPPH radical scavenging activity at different cocentrations of β-glucans compared with the universal reducing power (Vitamin C) showing means and standard deviation error bars from duplicate samples.*

### *3.4. Antioxidant activity of β-glucans polysaccharide*

As shown in Fig. 6, the percentage values of DPPH radical scavenging activity were increased gradually with an increase of the concentrations of β-glucans and the universal reducing power (vitamin C). The result also revealed that β-glucans exhibit more percent of scavenging activity (8.59, 12.36, 18.56, 23.69, 44.66, and 80.36%) than vitamin C (2.45, 4.75, 11.95, 19.45, 36.49, and 72.46%) at the concentrations of 31.2, 64.4, 125, 250, 500 and 800 μg/ml, respectively.

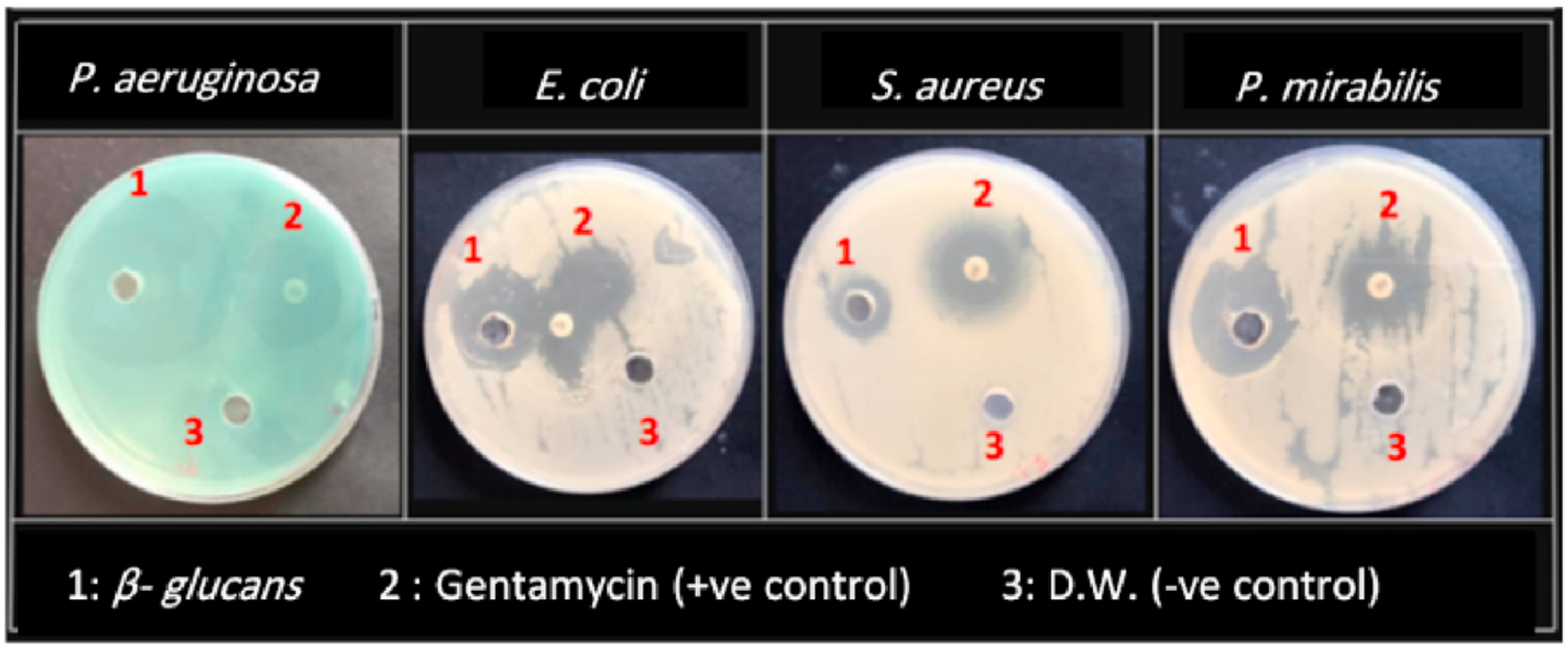

*Fig. 7. Antimicrobial activity of β-glucans against some pathogenic bacteria.*

*Table 1. The retention times of peaks of a polysaccharide β-glucans.*

| | Reten. Time | Result table (Uncal − F:\ β-glucans (10 ppm) − Detector 1) | | | | Compound Name |
|---|---|---|---|---|---|---|
| | | Area [mAU.s] | Height [mAU] | Area [%] | Height W 05 [min] | |
| 1 | 5.927 | 3299.544 | 208.935 | 100.0 | 100.0 0.25 | |
| | Total | 3299.544 | 208.935 | 100.0 | 100.0 | |

*Table 2. Result of FTIR spectroscopy frequency ranges and absorptions for functional groups.*

| S. No | Frequency | Assignment |
|---|---|---|
| 1 | 3344 | NH, OH |
| 2 | 3078 | = C−H |
| 3 | 2899 | C−H stretch |
| 4 | 1685 | C=O |
| 5 | 1564 | N−O, C−C |
| 6 | 1519 | C−C |
| 7 | 1413 | C−H |
| 8 | 1348 | N−O |
| 9 | 1066 | C−N |
| 10 | 819 | C−CL |
| 11 | 657 | C−Br |

Moreover, the standard deviations with β-glucans were (0.834386002, 0.509116882, 0.791959595, 0.975807358, 0.933380951 and 0.862670273), while the standard deviations with vitamin C were (0.636396103, 0.636396103, 0.777817459, 0.707106781, 0.721248917 and 0.763675324), at the above concentrations, respectively (Fig. 6).

*Table 3. The average numbers of antibacterial activity of β-glucans. Inhibition zone (Diameter in mm).*

| S. No | Microorganisms | Distilled water (-ve control) | Chloramphenicol mg/mL 40 (+ve control) | β-glucans 100 mg/mL |
|---|---|---|---|---|
| 1 | *P. aeruginosa* | 0 | 22 | 45 |
| 2 | *E. coli* | 0 | 21 | 23 |
| 3 | *S. aureus* | 0 | 20 | 13 |
| 4 | *P. mirabilis* | 0 | 24 | 24 |

*Table 4. Effect of purified β-glucans polysaccharide isolated from P. ostreatus on MCF-7 and WRL-68 lines using MTT assay for 24 h at 37 °C.*

| Cancer cell line | Concentrations (μg/ml) | | | | |
|---|---|---|---|---|---|
| | 1.0 | 1.5 | 2.0 | 2.5 | 3.0 |
| | Percentage of growth-inhibiting of cancer cells | | | | |
| MCF-7 | 18 | 24 | 50 | 59 | 62 |
| WRL-68 | 4 | 6 | 9 | 13 | 22 |

### 3.5. *Antimicrobial activity of β-glucans polysaccharide*

The result showed remarkable antimicrobial activity against *P. aeruginosa*, where the highest growth inhibition zone was (45 mm) at 100 mg/mL, followed by *P. mirabilis* (24 mm) and *E. coli* (23 mm). The lowest antimicrobial activity *of β-glucans* appeared against *S. aureus* (13 mm), as shown in Fig. 7 and Table 3.

### 3.6. *Cytotoxic activity of β-glucans polysaccharide*

The cytotoxicity results of β-glucans on MCF-7 and WRL-68 lines showed a variation in the cytotoxicity effect at all concentrations of purified β-glucans polysaccharide isolated from *P. ostreatus*. Remarkably, the MCF-7 line was more affected in increasing the percentage of growth-inhibiting than the WRL-68 line. Significantly, it was found that the cytotoxicity increases with increasing the glucose concentration, which is reflected in increasing the percentage of inhibition of cancer cells. The highest values were 62% and 22% at a concentration of 3.0 μg/ml in MCF-7 and WRL-68 lines, respectively. While the lowest values of the percentage of growth inhibition were 18% and 4% a concentration of 1.0 μg/ml in MCF-7 and WRL-68 lines, respectively (Table 4).

## 4. Discussion

The 657 bp fragment of ITS region (GenBank accession no. MW457626) of fruiting bodies isolated from wild *P. ostreatus* was confirmed (Fig. 2), and the homology searches were blasted against the NCBI database (Fig. 3). This region of nuclear ribosomal DNA has been widely used (as a common sequenced DNA region) for molecular genetic identification in various fungi [31]. The FTIR spectrum result revealed the presence of β-glycosidic bonds such as the presence of (NH, C–H, =C–H, C=O, and other groups) (Fig. 5 and Table 2). FTIR spectroscopy is considered an outstanding technique for structural analysis of polysaccharides [58,59]. This procedure can easily predict the anomeric arrangement and position of the glycosidic bonds in glucans. It detects the glucans in different raw materials especially in the crude with high molecular fractions. For instance, two essential kinds of glucans can be found in the fruiting bodies of mushrooms which are linear (1 → 3)-α-D and branched (1 → 3) (1 → 6)-β-D glucans [60]. Reactive oxygen species are necessary to initiate mutagenesis. The inhibitory effect resulting from antioxidant activities may lead to DNA mutagenesis produced by oxidative stress [61]. Also, oxidative stress has been considered to be one of the primary causal factors for different diseases and ageing, the antioxidant capacity of *β-glucans* was examined. The DPPH radical scavenging properties can be attributed to their hydrogen donating capacity. Our findings (Fig. 6) are consistent with previous results [62] who mentioned that antioxidant efficiencies by inhibitory concentration on DPPH showed a significant difference compared to universal antioxidants such as ascorbic acid. Moreover, the current result is agreed with [63] findings who reported that exopolysaccharides extracts from *G. lucidum* at 50, 100, 250, 500 and 1000 μg/ml revealed maximum antioxidant ability (82.30 ± 1.2%).

As natural antimicrobial compounds are important, mushrooms need these compounds as a source of antibiotics during their development. Industrial antibiotics and antimicrobial drugs perhaps are dangerous to human health and often can cause antibiotic resistance [64]. Therefore, using natural antimicrobial compounds is important [65]. Interestingly, β-glucans in fungal cell walls have been detected to have an antimicrobial impact [66]. Our results (Fig. 7 and Table 3) share similarities with [67] findings who mentioned that polysaccharides isolated from *Ganoderma* species have antibacterial effects against different types of bacteria *in vitro*. The current result is also concurred well with [6,18,62,66] who found that mushrooms have antimicrobial agents.

Cancers are mainly caused by the mutations in genetic material as a result of the changes that occur in the sequence of DNA, and these mutations can influence the structure as well as the function of the encoded proteins, then in genetic characteristics. It is found that 90% of the major reasons of cancers are correlated with mutations as a consequence of environmental factors for instance chemicals, radiation, metals, and others [68]. According to our results, it can reasonably assume that β-glucans polysaccharide helps to prevent or decrease the growth of cancer cells and maybe prevent the mutations as well. Our results (Table 4) are consistent with [48] who found that *Ganoderma lucidum* polysaccharide (GLP) isolated from *G. lucidum* which is the major compound of β-glucans can inhibit the mutations, also it is in complete agreement with [69] that β-glucans extracted from oyster mushrooms displays anti-neoplastic activity against MCF-7 line [70]. The reason for the anticancer properties is probably because of the activity of β-glucans polysaccharide, particularly in β-1,3 bound sites [71,72]. The current results are also in agreement with [18,73] who mentioned that bioactive molecule in *Pleurotus* sp. (oyster mushroom) involves high molecular weight compounds (polysaccharides) especially $\beta$-glucans which has anti-cancer effects.

## 5. Conclusion

To sum up, the ITS region of *P. ostreatus* was molecularly confirmed. Multiple sequence alignment and phylogenetic tree of studied ITS were performed depending on some available sequences in GenBank. The analysis of HPLC and FTIR has revealed the presence of $\beta$-glucans polysaccharide in isolated specimens. The current result revealed the highest antioxidant activity of $\beta$-glucans was 80.36% at 800 μg/ml. Moreover, a larger zone of inhibition of bacterial growth was (45 mm) against *P. aeruginosa* at 100 mg/ml. The highest percentage of growth inhibition in MCF-7 line was 62% at 3.0 μg/ml comparing with 22% in WRL-68 line. It seems that $\beta$-glucans polysaccharide of *P. ostreatus* has various health benefits and therapeutic properties; and it can be considered safe alternatives for human food and health.

## Acknowledgement

This work was performed at the research unit of the Department of the Biology/University of Mosul. The authors would like to thank the Research Center of the Biotechnology at the University of Al-Nahrain for their kind assistance to provide MCF-7 and WRL-68 cancer cell lines.